\documentclass[conference]{IEEEtran}
\IEEEoverridecommandlockouts

\usepackage{colortbl}
\usepackage{cite}
\usepackage{amsmath,amssymb,amsfonts}
\usepackage{algorithmic}
\usepackage{graphicx}
\usepackage{textcomp}
\usepackage{pgfplots}
\pgfplotsset{compat=newest}
\usepgfplotslibrary{statistics}
\usepackage{array}
\usepackage{multirow}
\usepackage{booktabs}
\usepackage{adjustbox}
\usepackage{verbatim}
\usepackage{url}
\usepackage{subcaption}
\usepackage{float}
\usepackage{tikz}
\usetikzlibrary{patterns, shapes, arrows, positioning}
\definecolor{cellblue}{rgb}{0.7, 0.8, 0.9}
\definecolor{cellbluestrong}{rgb}{0.2, 0.3, 0.5}
\definecolor{lightblue}{rgb}{0.8,0.85,1}
\definecolor{lightgreen}{rgb}{0.85,1,0.85}

\def\BibTeX{{\rm B\kern-.05em{\sc i\kern-.025em b}\kern-.08em
    T\kern-.1667em\lower.7ex\hbox{E}\kern-.125emX}}

\begin{document}


\title{\textit{Are Music Foundation Models Better at Singing Voice Deepfake Detection?} Far-Better Fuse them with Speech Foundation Models}

\author{
    \IEEEauthorblockN{Orchid Chetia Phukan\textsuperscript{*1}, Sarthak Jain\textsuperscript{*1}, Swarup Ranjan Behera\textsuperscript{2}\\ Arun Balaji Buduru\textsuperscript{1}, Rajesh Sharma\textsuperscript{1,3}, S.R Mahadeva Prasanna\textsuperscript{4,5}}
    \IEEEauthorblockA{
        $^1$\textit{IIIT-Delhi, India},
        $^2$\textit{Reliance Jio AICoE, India},
        $^3$\textit{University of Tartu, Estonia} \\
        $^4$\textit{IIT-Dharwad, India},
        $^5$\textit{IIIT-Dharwad, India}\\
        *equal contribution\\
        orchidp@iiitd.ac.in
    }
}

\maketitle
 
\begin{abstract}
     In this study, for the first time, we extensively investigate whether music foundation models (MFMs) or speech foundation models (SFMs) work better for singing voice deepfake detection (SVDD), which has recently attracted attention in the research community. For this, we perform a comprehensive comparative study of state-of-the-art (SOTA) MFMs (MERT variants and music2vec) and SFMs (pre-trained for general speech representation learning as well as speaker recognition). We show that speaker recognition SFM representations perform the best amongst all the foundation models (FMs), and this performance can be attributed to its higher efficacy in capturing the pitch, tone, intensity, etc, characteristics present in singing voices. To our end, we also explore the fusion of FMs for exploiting their complementary behavior for improved SVDD, and we propose a novel framework, \textbf{FIONA} for the same. With \textbf{FIONA}, through the synchronization of x-vector (speaker recognition SFM) and MERT-v1-330M (MFM), we report the best performance with the lowest Equal Error Rate (EER) of 13.74 \%, beating all the individual FMs as well as baseline FM fusions and achieving SOTA results. 
\end{abstract}


\section{Introduction}
    \begin{quote}
        ``An A.I. Hit of Fake ‘Drake’ and ‘The Weeknd’ Rattles the Music World.''
        \begin{flushright}
            --- \textit{The New York Times}~\cite{nytimes2024}
        \end{flushright}
    \end{quote}
    
    This headline encapsulates the profound disruption that AI-generated content is causing in the music industry. With the rapid advancements in singing voice synthesis (SVS)~\cite{svs1,svs2,svs3} and singing voice conversion (SVC)~\cite{svc1,svc2,svc3,svc4}, models like VISinger~\cite{vis} and DiffSinger~\cite{diff} can now produce synthetic vocals that closely mimic real artists with remarkable accuracy. These technologies not only replicate the pitch, timbre, and emotional depth of a singer's voice but also threaten to undermine the commercial value and intellectual property of original artists. The ease of creating high-fidelity deepfake singing voices has raised alarms among artists, record labels, and publishers, leading to a critical need for robust detection methods. 

    Despite significant advances in the broader domain of audio deepfake detection (ADD)~\cite{ADD1,ADD2}, most research has primarily focused on detecting fake speech rather than singing voices. ADD has evolved from using handcrafted statistical features to leveraging modern deep learning algorithms, including the usage of foundation models (FMs), pre-trained on large amounts of data. These approaches have achieved impressive results in speech deepfake detection, with some systems~\cite{eer1,eer2,eer3} reporting EERs below 1\% on benchmark datasets like ASVspoof2019~\cite{ASVspoof2A}. However, recent studies, including work by Zang et al.~\cite{SingFakeSV}, highlight that SOTA speech ADD systems often perform poorly on singing voices, showing significant degradation. These findings 
    underscores the critical need for dedicated SVDD systems tailored to the unique characteristics of singing voices. 

    In the context of SVDD, the specific demands of accurately detecting fake singing voices require a deeper exploration of the suitability of various FMs. The FMs are preferred due to the performance benefit as well as preventing training models from scratch. Music Foundation Models (MFMs) are tailored to understand musical content, rhythmic understanding, and timbre but may lack the nuanced speech features crucial for detecting subtle alterations in vocal authenticity. Conversely, Speech Foundation Models (SFMs) have demonstrated superior performance in capturing the intricacies of human speech, including pitch, tone, and intensity; their effectiveness in the musical domain, particularly for singing voices, is less certain. Given this, our study investigates the comparative performance of these FMs for SVDD. Furthermore, inspired by research on speech deefake detection that combines FMs for better performance by exploiting their complementary behavior~\cite{ADD2}, we also investigate this for SVDD. Chen et al.
 \cite{chen24o_interspeech} took an initial step towards this direction, however, experimenting only with the combination wav2vec2 and MERT-v1-330M and not exploring other SOTA MFMs and SFMs. In our study, we address this shortcoming of previous research. 

    In this paper, we make the following contributions:
    
    \begin{itemize}
        \item We conduct a detailed comparison of SOTA MFMs and SFMs for SVDD on the benchmark CtrSVDD dataset, evaluating their ability to capture the distinctive features of fake singing voices.
        \item We introduce a novel framework, \textbf{FIONA} (\textbf{F}us\textbf{ION} through Kernel \textbf{A}lignment), that synergistically combines the strengths of different FMs. \textbf{FIONA} achieves SOTA results in SVDD by aligning representations from x-vector (Speaker recognition SFM) and MERT-v1-330M (MFM), significantly outperforming individual FMs as well as baseline fusion technique. 
    \end{itemize}

    \noindent We will release the models and codes created as part of this work after the review cycle. 
    
    
\section{Foundation Models}
    In this section, we discuss the SOTA SFMs and MFMs considered in our study. We select these FMs due to their exceptional performance in their respective benchmarks.

    \subsection{Speech Foundation Models (SFMs)}
    
        \noindent \textbf{Unispeech-SAT\footnote{\url{https://huggingface.co/microsoft/unispeech-sat-base}} ~\cite{chen2022unispeech}}: It is a contrastive loss-based multitask learning model with speaker-aware pre-training, pre-trained on 94k hours of VoxPopuli, LibriVox, and Gigaspeech datasets.
        
        \noindent \textbf{WavLM\footnote{\url{https://huggingface.co/microsoft/wavlm-base}}~\cite{chen2022wavlm}}: Trained in a self-supervised manner for both speech-masked modeling and speech denoising. It is a pre-trained as general-purpose representation learning model and shows SOTA performance across various speech processing tasks.
        
        \noindent \textbf{Wav2vec2\footnote{\url{https://huggingface.co/facebook/wav2vec2-base}}~\cite{baevski2020wav2vec}}: It is not a SOTA FM like Unispeech-SAT and WavLM. However, we include them in our study due to its effectiveness in related tasks such as speech emotion recognition ~\cite{pepino21_interspeech} and speech deepfake detection ~\cite{kang2024experimental}. 
        
        \noindent \textbf{x-vector\footnote{\url{https://huggingface.co/speechbrain/spkrec-xvect-voxceleb}}~\cite{8461375}}: It is a time-delay neural network and shows SOTA performance for speaker recognition. We include x-vector in our experiments due to its effectiveness in related tasks such as speech emotion recognition~\cite{Phukan2023TransformingTE} and shout intensity prediction~\cite{fukumori2023investigating}. 
        
    
    \subsection{Music Foundation Models (MFMs)}
    
        \noindent \textbf{MERT Series~\cite{li2023mert}:} The MERT (Music Embedding Representation Transformer) series represents SOTA MFMs, designed to capture the intricate features of musical audio. These models excel in tasks such as genre classification and mood prediction, underpinned by robust pretraining on diverse musical datasets. The following MERT variants were considered for this study: \textit{MERT-v1-330M}\footnote{\url{https://huggingface.co/m-a-p/MERT-v1-330M}}, \textit{MERT-v1-95M}\footnote{\url{https://huggingface.co/m-a-p/MERT-v1-95M}}, \textit{MERT-v0-public}\footnote{\url{https://huggingface.co/m-a-p/MERT-v0-public}}, and \textit{MERT-v0}\footnote{\url{https://huggingface.co/m-a-p/MERT-v0}}.
        
        
        \noindent \textbf{music2vec-v1\footnote{\url{https://huggingface.co/m-a-p/music2vec-v1}}~\cite{Music2VecAS}}: It is trained in a self-supervised fashion. It excels in capturing the nuances of musical attributes and has been widely used in various downstream music tasks. 
    
    All MFMs require audio at specific sampling rates: MERT-v1-330M and MERT-v1-95M at 24 kHz, and MERT-v0-public, MERT-v0, music2vec-v1 at 16 kHz. The SFMs also require the audio to be resampled to 16Khz. Therefore, we resample the audio to these required rates before processing it with each FM. Representations are extracted by applying average pooling to the last hidden layer of FM. The dimension size of representations from music2vec-v1 and MERT-variants except MERT-v1-330M is 768 as well as for all the SFMs except x-vector. For MERT-v1-330M and x-vector, the dimension size is 1024 and 512 respectively. 

    \begin{figure}[bt] 
        \centering
        \includegraphics[width=0.475\textwidth]{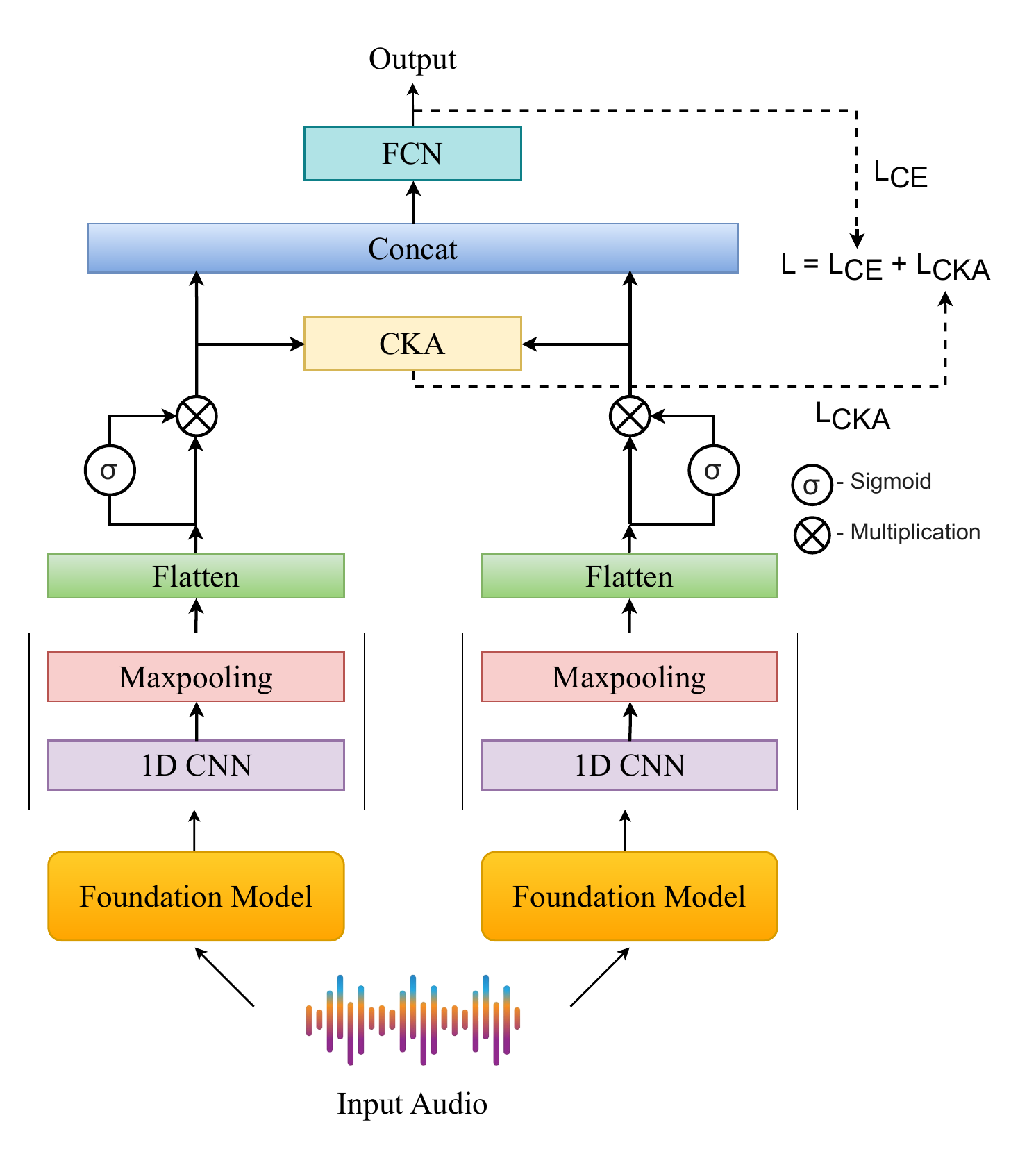} 
        \caption{\textbf{FIONA} framework: CKA and FCN stand for centered kernel alignment and fully connected network, respectively; $L$, $L_{CE}$, and $L_{CKA}$ represent the total loss, cross-entropy loss, and centered kernel alignment loss, respectively.}
        \label{fig:archi_cka}
    \end{figure}
\vspace{-0.2cm}
\section{Modeling}
    In this section, we discuss the downstream networks used for individual FMs as well as the proposed framework, \textbf{FIONA} for the fusion of FMs. 
     \vspace{-0.2cm}
    \subsection{Individual Representations Modeling}
    We use a fully connected network (FCN) and CNN on top of the extracted representations from different FMs. FCN is structured with dense layers of 128, 64, and 32 neurons. The CNN architecture includes two 1D-CNN layers with 16 and 32 filters with 3x3 size, each followed by max-pooling. We add a dense layer with 50 neurons and an output layer of 2 neurons, representing the non-fake and fake classes. 
    \vspace{-0.2cm}
    \subsection{Fusion through Concatenation}
    We use the same modeling and architectural details as in individual representations modeling and as a fusion mechanism, we use simple concatenation. Architecture is the same as in Figure \ref{fig:archi_cka}, except for the CKA part and the gating part. We fuse the features from the individual FM network after the flattening layer. The fused features are passed through a dense layer with 50 neurons followed by an output layer of 2 neurons. We use cross-entropy as the loss function and Adam as the optimizer in both the individual representations modeling experiments as well as with the fusion through concatenation experiments.
    \vspace{-0.5cm}
    \subsection{Fusion through Kernel Alignment (FIONA)}
        The architecture of the proposed framework, \textbf{FIONA} is shown in Figure \ref{fig:archi_cka}. We follow the same architectural design as the fusion through concatenation; however, in \textbf{FIONA}, we introduce a gating mechanism after flattening the features from FMs with a sigmoid function that makes sure only essential features are passed to the next stage. We also linearly project the features of both FM networks. To effectively fuse features from different FMs, we introduce Centered Kernel Alignment (CKA) as a novel loss function. Traditionally, CKA is used for measuring representational similarity~\cite{CKA1,CKA2,CKA3,CKA4,CKA5}. However, we repurpose CKA as a loss function in our framework to drive effective alignment and integration between FM features. This innovative application of CKA, embedded within \textbf{FIONA}, leverages the complementary strengths of different FMs for better SVDD.

    \noindent \textbf{CKA as a Loss Function:} 
    
    
    \noindent Let $\mathbf{X} \in \mathbb{R}^{n \times d}$ and $\mathbf{Y} \in \mathbb{R}^{n \times d}$ denote two feature matrices from two FMs networks, where \(n\) is the number of samples and \(d\) is the feature dimension. The CKA between $\mathbf{X}$ and $\mathbf{Y}$ is computed as follows:
    
    \noindent \textbf{1. Gram Matrix Calculation:}
    Computing the Gram matrices for the feature sets $\mathbf{X}$ and $\mathbf{Y}$:
    \begin{equation}
    \mathbf{K} = \mathbf{X}\mathbf{X}^\top
    \end{equation}
    \begin{equation}
    \mathbf{L} = \mathbf{Y}\mathbf{Y}^\top
    \end{equation}
    
    \noindent \textbf{2. Centering the Gram Matrices:}
    Centering these matrices using the centering matrix $\mathbf{H} = \mathbf{I} - \frac{1}{n}\mathbf{1}\mathbf{1}^\top$, where $\mathbf{I}$ is the identity matrix and $\mathbf{1}$ is a vector of ones:
    \begin{equation}
    \tilde{\mathbf{K}} = \mathbf{H}\mathbf{K}\mathbf{H}
    \end{equation}
    \begin{equation}
    \tilde{\mathbf{L}} = \mathbf{H}\mathbf{L}\mathbf{H}
    \end{equation}
    
    \noindent \textbf{3. Hilbert-Schmidt Independence Criterion (HSIC):}
    Calculating the HSIC between the centered Gram matrices $\tilde{\mathbf{K}}$ and $\tilde{\mathbf{L}}$:
    \begin{equation}
    \text{HSIC}(\tilde{\mathbf{K}}, \tilde{\mathbf{L}}) = \text{trace}(\tilde{\mathbf{K}}\tilde{\mathbf{L}})
    \end{equation}
    
    \noindent \textbf{4. CKA Computation:}
    Finally, computing the CKA as:
    \begin{equation}
    \text{CKA}(\mathbf{X}, \mathbf{Y}) = \frac{\text{HSIC}(\tilde{\mathbf{K}}, \tilde{\mathbf{L}})}{\sqrt{\text{HSIC}(\tilde{\mathbf{K}}, \tilde{\mathbf{K}}) \cdot \text{HSIC}(\tilde{\mathbf{L}}, \tilde{\mathbf{L}})}}
    \end{equation}
    Here, $\text{HSIC}(\tilde{\mathbf{K}}, \tilde{\mathbf{L}})$ measures the independence between the centered Gram matrices, while $\text{HSIC}(\tilde{\mathbf{K}}, \tilde{\mathbf{K}})$ and $\text{HSIC}(\tilde{\mathbf{L}}, \tilde{\mathbf{L}})$ normalize the alignment score. \par

    \noindent \textbf{Total Loss Function:}
    \noindent Here, CKA as a loss function reduces the distance between the features of the FMs and aligns them in a joint distribution. Finally, we combine the CKA loss, $\mathcal{L}_{CKA}$ with cross-entropy loss, $\mathcal{L}_{CE}$, and perform joint optimization. The cross-entropy loss is computed as:
    \begin{equation}
    \mathcal{L}_{CE} = -\frac{1}{n} \sum_{i=1}^n \sum_{c=1}^C y_{true}^{(i,c)} \log y_{pred}^{(i,c)}
    \end{equation}
    where $C$ represents the number of classes and $n$ represents the number of samples.
    The total  loss function is given by:
    \begin{equation}
    \mathcal{L} = \mathcal{L}_{CE} + \lambda \mathcal{L}_{CKA}
    \end{equation}
    where $\lambda$ is a hyperparameter that scales the contribution of the CKA loss term. 
    This approach not only minimizes classification error but also aligns feature representations, thereby enhancing fusion across different FMs.

\vspace{-0.2cm}
\section{Experiments}
\vspace{-0.2cm}
    \subsection{Benchmark Dataset}
    We utilize the CtrSVDD~\cite{CtrSVDDAB} dataset due to its extensive and diverse coverage. In contrast to previous datasets like SingFake~\cite{SingFakeSV}, which suffers from incomplete method disclosures and limited diversity, and FSD~\cite{FSDAI}, which is limited by licensing restrictions and a narrow range of deepfake methods, CtrSVDD offers 47.64 hours of bonafide and 260.34 hours of deepfake singing vocals, generated through 14 distinct methods involving 164 singer identities. This comprehensive dataset enables a more robust evaluation of SVDD models. 
    
\vspace{-0.2cm}
    \subsection{Training Details}
    We train all models for 50 epochs with a learning rate of $1 \times 10^{-3}$ and batch size of 32. We use dropout and early stopping techniques to mitigate overfitting and use Adam as the optimizer. We train all the models in the training set of CtrSVDD and test it on the development set of CtrSVDD, as labels were not provided for the test set. The training parameters of the models for individual FM representations range from 0.6M to 1.1M. For \textbf{FIONA}, with different FMs combinations range from 2.6M to 3.1M trainable parameters.

    \begin{table}[bt] 
    \centering
    \begin{tabular}{c|c|c}
        \toprule
        {\textbf{FM}} & {\textbf{FCN}} & {\textbf{CNN}} \\ \midrule
         {MFCC} & 43.99 & 41.18 \\
        {x-vector} & \cellcolor{blue!25}\textbf{17.35} & \cellcolor{blue!25}\textbf{14.18}\\
        {Unispeech-SAT} & \cellcolor{green!25}\textbf{22.78} & \cellcolor{yellow!25}\textbf{22.72} \\
        {WavLM} & 37.48 & 32.12 \\
        {Wav2vec2} & \cellcolor{yellow!25}\textbf{24.02} & \cellcolor{green!25}\textbf{18.94} \\
        {MERT-v1-330M} & 29.46 & 27.50 \\
        {MERT-v1-95M} & 35.19 & 33.01 \\
        {MERT-v0-public} & 43.97 & 37.91 \\
        {MERT-v0} & 37.80 & 36.41 \\
        {music2vec-v1} & 50.00 & 35.81 \\
        \bottomrule
    \end{tabular}
    \caption{Equal Error Rate (EER) scores for different FMs using CNN and FCN as downstream models: All scores are in \%; lower EER is better. We use MFCC features as a baseline.}
    \label{tab:single_ptms_table}
    \end{table}

\vspace{-0.2cm}
    \subsection{Experimental Results}
\begin{figure}[bt]
    \centering
    \begin{minipage}{0.22\textwidth}
        \centering
        \includegraphics[width=1.05\textwidth]{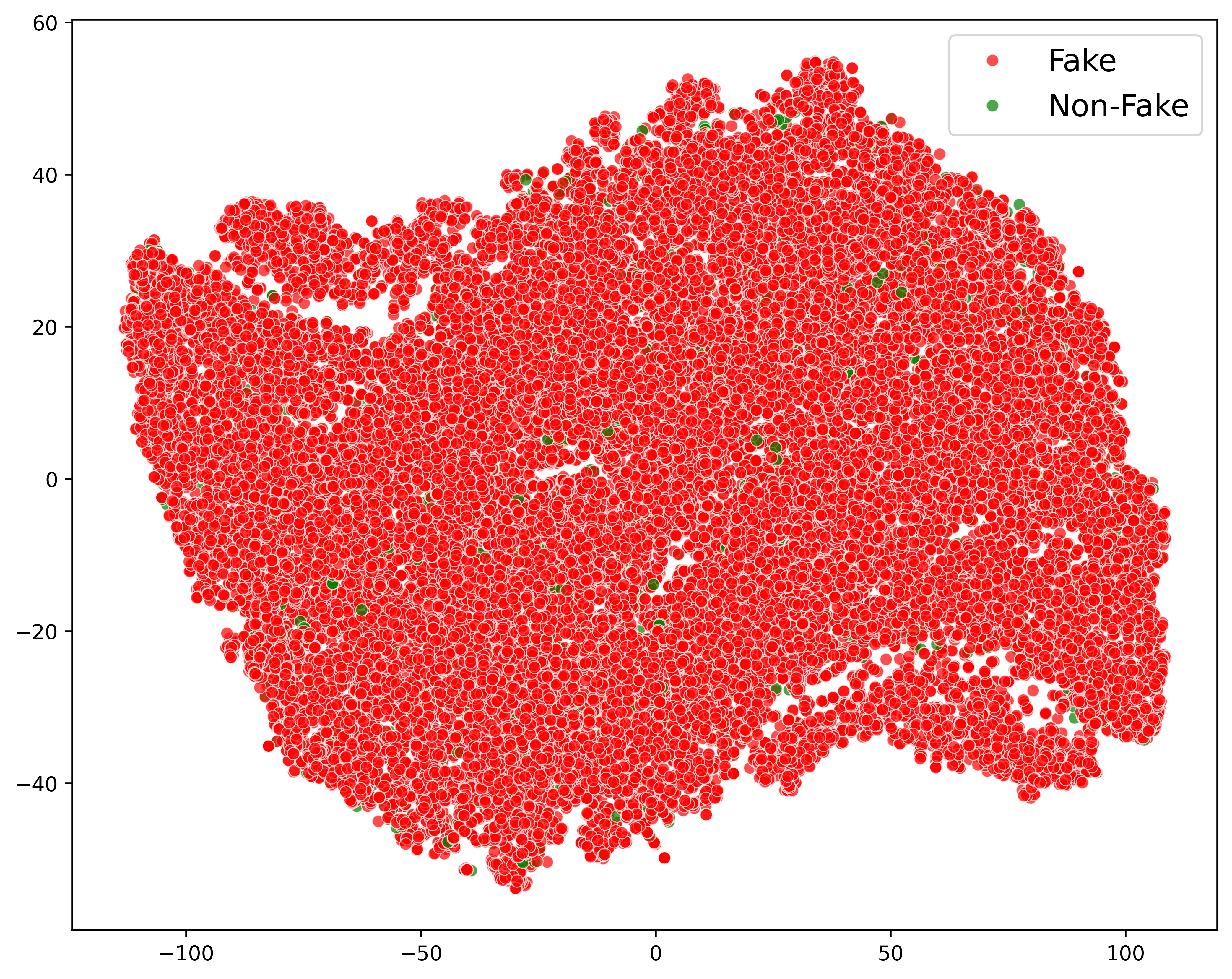} 
        \caption*{MERT\_v1\_330M}
    \end{minipage}\hfill \hfill
    \begin{minipage}{0.22\textwidth}
        \centering
        \includegraphics[width=1.05\textwidth]{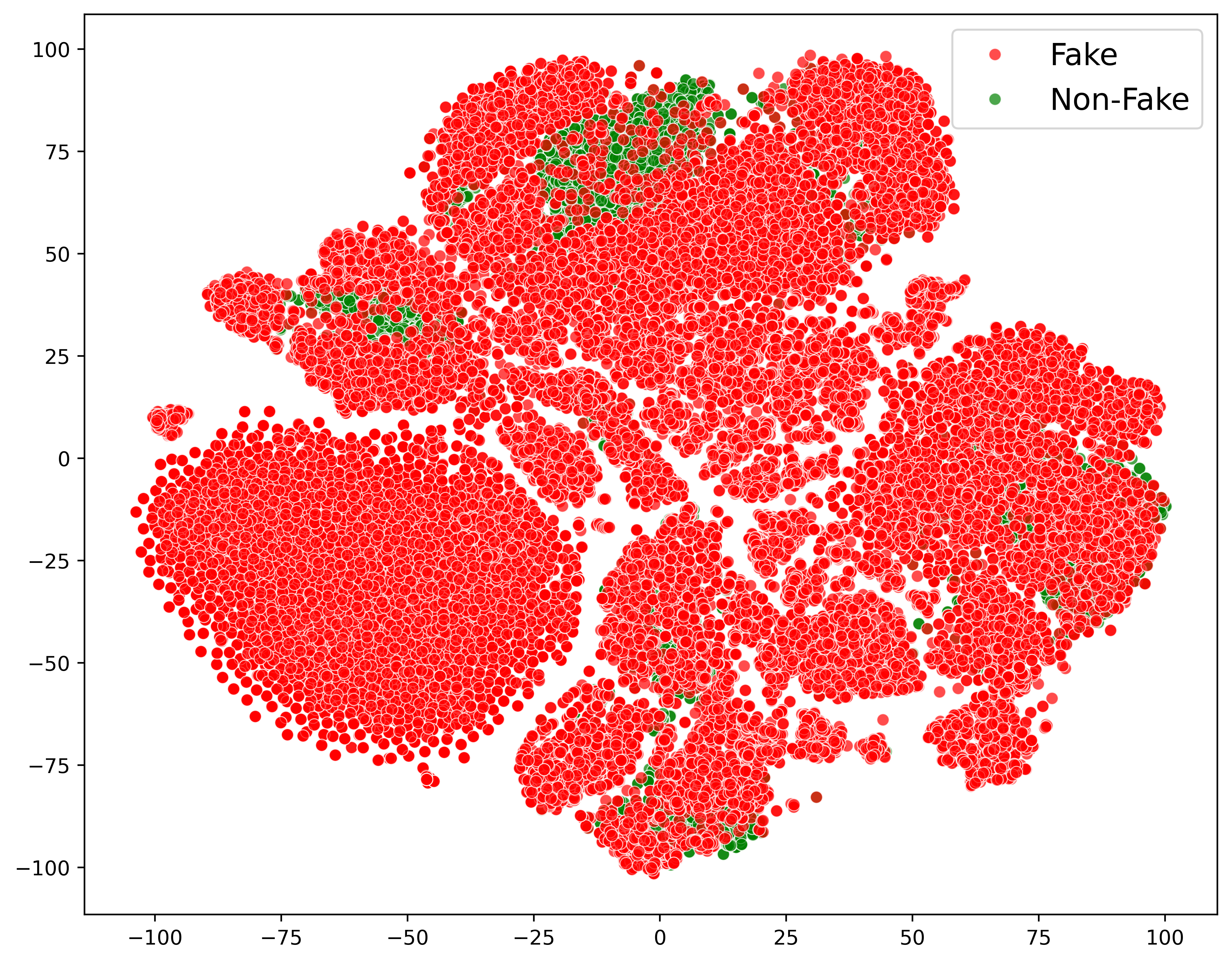} 
        \caption*{x-vector}
    \end{minipage}\\[10pt] 
    \caption{t-SNE plots of raw FM representations.}
    \label{fig:tsne}
\end{figure}

    Table~\ref{tab:single_ptms_table} shows the Equal Error Rate (EER) scores for FCN and CNN models across various FMs representations. x-vector, speaker recognition SFM representations achieve the lowest EER of 14.18\% with CNN and 17.35 \% with FCN amongst all the FMs, including both SFMs and MFMs, indicating superior performance for SVDD. This can be attributed to its effectiveness in capturing the pitch, tone, intensity, of the singing voices for better SVDD. In summary, we see that SFMs show low EER in contrast to MFMs, thus showing that accurately detecting speech characteristics leads to better SVDD performance. Among the MFMs, MERT-v1-330M representations showed the topmost performance, which can be due to its larger size. Among the SFMs, Unispeech-SAT and Wav2vec2, both showed relatively mixed performance, while WavLM performed the worst. We also plot the t-SNE plots of raw representations in Figure \ref{fig:tsne}, here also, we can observe better clustering across classes for x-vector, and thus supporting our results. 

\begin{table}[bt] 
\centering
{\fontsize{8}{10}\selectfont 
\resizebox{\columnwidth}{!}{
\begin{tabular}{c|c|c}
    \toprule
    \textbf{FM combinations} & \textbf{Concatenation} & \textbf{FIONA} \\
    \midrule
    x-vector + Unispeech-SAT & \cellcolor{green!25}\textbf{14.75} & \cellcolor{green!25}\textbf{13.90} \\
    x-vector + WavLM & \cellcolor{yellow!25}\textbf{15.64} & 15.90 \\
    x-vector + Wav2vec2 & 16.12 & \cellcolor{yellow!25}\textbf{14.90} \\
    x-vector + MFCC & 16.06 & 14.91 \\
    Unispeech-SAT + WavLM & 22.88 & 20.18 \\
    Unispeech-SAT + Wav2vec2 & 20.19 & 17.74 \\
    Unispeech-SAT + MFCC & 37.99  & 19.85 \\
    WavLM + Wav2vec2 & 22.39 & 21.13 \\
    WavLM + MFCC & 36.16  & 35.87\\
    Wav2vec2 + MFCC & 28.16 & 23.94 \\
    x-vector + MERT-v1-330M & \cellcolor{blue!25}\textbf{13.85} & \cellcolor{blue!25}\textbf{13.74} \\
    x-vector + MERT-v1-95M & 20.16& 16.30  \\
    x-vector + MERT-v0-public & 20.05 &  17.68\\
    x-vector + MERT-v0 & 20.01& 16.37\\
    x-vector + music2vec-v1 & 20.27& 16.75 \\
    Unispeech-SAT + MERT-v1-330M & 20.65 & 19.25\\
    Unispeech-SAT + MERT-v1-95M & 21.38 & 20.45 \\
    Unispeech-SAT + MERT-v0-public & 22.33 & 19.05 \\
    Unispeech-SAT + MERT-v0 & 22.17 & 20.07 \\
    Unispeech-SAT + music2vec-v1 & 22.27 & 20.89 \\
    WavLM + MERT-v1-330M & 28.24 & 21.06 \\
    WavLM + MERT-v1-95M & 30.11 & 23.05 \\
    WavLM + MERT-v0-public & 31.18 & 30.25 \\
    WavLM + MERT-v0 & 34.95 & 28.26 \\
    WavLM + music2vec-v1 & 32.00 & 29.74 \\
    Wav2vec2 + MERT-v1-330M & 25.07 & 22.35\\
    Wav2vec2 + MERT-v1-95M & 22.82 & 20.97 \\
    Wav2vec2 + MERT-v0-public & 24.06 & 22.55 \\
    Wav2vec2 + MERT-v0 & 23.06 & 20.23 \\
    Wav2vec2 + music2vec-v1 & 23.74 & 20.91 \\
    MFCC + MERT-v1-330M & 37.16 & 32.26 \\
    MFCC + MERT-v1-95M & 36.84 & 36.50 \\
    MFCC + MERT-v0`-public & 40.23 & 32.05 \\
    MFCC + MERT-v0 & 41.59 & 40.04 \\
    MFCC + music2vec-v1 & 39.86 & 37.06 \\
    MERT-v1-330M + MERT-v1-95M & 29.83 & 22.13 \\
    MERT-v1-330M + MERT-v0-public & 29.98 & 22.00 \\
    MERT-v1-330M + MERT-v0 & 30.19 & 20.24 \\
    MERT-v1-330M + music2vec-v1 & 26.65 & 21.33 \\
    MERT-v1-95M + MERT-v0-public & 34.42 & 22.98 \\
    MERT-v1-95M + MERT-v0 & 33.53 & 23.19 \\
    MERT-v1-95M + music2vec-v1 & 50.00 & 23.01 \\
    MERT-v0-public + MERT-v0 & 39.72 & 20.37 \\
    MERT-v0-public + music2vec-v1 & 39.96 & 21.87 \\
    MERT-v0 + music2vec-v1 & 34.65 & 22.79 \\
    \bottomrule
\end{tabular}
}
}
\caption{Equal Error Rate (EER) scores for concatenation-based fusion and fusion using \textbf{FIONA}: All scores are given in \%; lower EER indicates better performance.} 
\label{tab:combination_ptms_table}
\end{table}

    Table~\ref{tab:combination_ptms_table} presents the EER scores for fusion of FMs with baseline concatenation-based fusion and fusion with \textbf{FIONA}. Here, we only experimented with CNN as the downstream network due to its top performance with individual FM representations in Table \ref{tab:single_ptms_table}. Notably, \textbf{FIONA} with x-vector + MERT-v1-330M combination with achieves the lowest EER of 13.74\% amongst all the combinations of FMs as well as baseline concatenation-based fusion, thus showing the efficacy of \textbf{FIONA} for effective fusion of FMs. Fusion with \textbf{FIONA}, generally leads to better results than concatenation-based fusion. We also observe that the fusion of MFMs with each other generally leads to lower performance, thus showing conflicting behavior, however, in contrast we observe better complementary behavior amongst SFMs. Furthermore, we observe that the fusion of the x-vector with other SFMs and MFMs generally leads to better results and thus shows observable complementary behavior. 
    

\vspace{-0.3cm}
\section{Conclusion}
In conclusion, our study presents the first comprehensive investigation into whether MFMs or SFMs are more effective for SVDD, an area with increased traction. Our extensive comparative analysis of SOTA MFMs (MERT variants and music2vec) and SFMs (pre-trained for general speech representation and speaker recognition) reveals that speaker recognition SFMs outperform all other FMs, likely due to their superior ability to capture key singing voice characteristics like pitch, tone, and intensity. Additionally, we introduce a novel fusion framework, \textbf{FIONA}, for leveraging the complementary strengths of different FMs for enhanced SVDD. By synchronizing x-vector (speaker recognition SFM) with MERT-v1-330M (MFM), \textbf{FIONA} achieves the best performance, with the lowest EER of 13.74\%, surpassing individual FMs and baseline fusion technique, and setting a new SOTA for SVDD. Our work will act as a guide for future researchers for appropriate selections of FM representations for SVDD and also as an inspiration for coming up with more effective fusion techniques for combining FMs. 


\bibliographystyle{IEEEtran}
\bibliography{mybib}

\begin{thebibliography}{10}
\providecommand{\url}[1]{#1}
\csname url@samestyle\endcsname
\providecommand{\newblock}{\relax}
\providecommand{\bibinfo}[2]{#2}
\providecommand{\BIBentrySTDinterwordspacing}{\spaceskip=0pt\relax}
\providecommand{\BIBentryALTinterwordstretchfactor}{4}
\providecommand{\BIBentryALTinterwordspacing}{\spaceskip=\fontdimen2\font plus
\BIBentryALTinterwordstretchfactor\fontdimen3\font minus \fontdimen4\font\relax}
\providecommand{\BIBforeignlanguage}[2]{{%
\expandafter\ifx\csname l@#1\endcsname\relax
\typeout{** WARNING: IEEEtran.bst: No hyphenation pattern has been}%
\typeout{** loaded for the language `#1'. Using the pattern for}%
\typeout{** the default language instead.}%
\else
\language=\csname l@#1\endcsname
\fi
#2}}
\providecommand{\BIBdecl}{\relax}
\BIBdecl

\bibitem{nytimes2024}
\BIBentryALTinterwordspacing
T.~N.~Y. Times, ``An a.i. hit of fake ‘drake’ and ‘the weeknd’ rattles the music world,'' 2024, accessed: 2024-08-20. [Online]. Available: \url{https://www.nytimes.com/2023/04/19/arts/music/ai-drake-the-weeknd-fake.html}
\BIBentrySTDinterwordspacing

\bibitem{svs1}
P.~Lu, J.~Wu, J.~Luan, X.~Tan, and L.~Zhou, ``Xiaoicesing: A high-quality and integrated singing voice synthesis system,'' \emph{ArXiv}, vol. abs/2006.06261, 2020.

\bibitem{svs2}
J.~Chen, X.~Tan, J.~Luan, T.~Qin, and T.-Y. Liu, ``Hifisinger: Towards high-fidelity neural singing voice synthesis,'' \emph{ArXiv}, vol. abs/2009.01776, 2020.

\bibitem{svs3}
Y.~Ren, X.~Tan, T.~Qin, J.~Luan, Z.~Zhao, and T.-Y. Liu, ``Deepsinger: Singing voice synthesis with data mined from the web,'' \emph{Proceedings of the 26th ACM SIGKDD International Conference on Knowledge Discovery \& Data Mining}, 2020.

\bibitem{svc1}
W.-C. Huang, L.~P. Violeta, S.~Liu, J.~Shi, Y.~Yasuda, and T.~Toda, ``The singing voice conversion challenge 2023,'' \emph{2023 IEEE Automatic Speech Recognition and Understanding Workshop (ASRU)}, pp. 1--8, 2023.

\bibitem{svc2}
T.~Jayashankar, J.~Wu, L.~Sari, D.~Kant, V.~Manohar, and Q.~He, ``Self-supervised representations for singing voice conversion,'' \emph{ICASSP 2023 - 2023 IEEE International Conference on Acoustics, Speech and Signal Processing (ICASSP)}, pp. 1--5, 2023.

\bibitem{svc3}
A.~Polyak, L.~Wolf, Y.~Adi, and Y.~Taigman, ``Unsupervised cross-domain singing voice conversion,'' in \emph{Interspeech}, 2020.

\bibitem{svc4}
C.~Deng, C.~Yu, H.~Lu, C.~Weng, and D.~Yu, ``Pitchnet: Unsupervised singing voice conversion with pitch adversarial network,'' \emph{ICASSP 2020 - 2020 IEEE International Conference on Acoustics, Speech and Signal Processing (ICASSP)}, pp. 7749--7753, 2019.

\bibitem{vis}
Y.~Zhang, J.~Cong, H.~Xue, L.~Xie, P.~Zhu, and M.~Bi, ``Visinger: Variational inference with adversarial learning for end-to-end singing voice synthesis,'' \emph{ICASSP 2022 - 2022 IEEE International Conference on Acoustics, Speech and Signal Processing (ICASSP)}, pp. 7237--7241, 2021.

\bibitem{diff}
J.~Liu, C.~Li, Y.~Ren, F.~Chen, and Z.~Zhao, ``Diffsinger: Singing voice synthesis via shallow diffusion mechanism,'' in \emph{AAAI Conference on Artificial Intelligence}, 2021.

\bibitem{ADD1}
J.~Yi, C.~Wang, J.~Tao, X.~Zhang, C.~Y. Zhang, and Y.~Zhao, ``Audio deepfake detection: A survey,'' \emph{ArXiv}, vol. abs/2308.14970, 2023.

\bibitem{ADD2}
O.~C. Phukan, G.~S. Kashyap, A.~B. Buduru, and R.~Sharma, ``Heterogeneity over homogeneity: Investigating multilingual speech pre-trained models for detecting audio deepfake,'' \emph{Findings of the Association for Computational Linguistics: NAACL 2024}, 2024.

\bibitem{eer1}
J.~weon Jung, H.-S. Heo, H.~Tak, H.~jin Shim, J.~S. Chung, B.-J. Lee, H.~jin Yu, and N.~W.~D. Evans, ``Aasist: Audio anti-spoofing using integrated spectro-temporal graph attention networks,'' \emph{ICASSP 2022 - 2022 IEEE International Conference on Acoustics, Speech and Signal Processing (ICASSP)}, pp. 6367--6371, 2021.

\bibitem{eer2}
J.~Xue, C.~Fan, J.~Yi, C.~Wang, Z.~Wen, D.~Zhang, and Z.~Lv, ``Learning from yourself: A self-distillation method for fake speech detection,'' \emph{ICASSP 2023 - 2023 IEEE International Conference on Acoustics, Speech and Signal Processing (ICASSP)}, pp. 1--5, 2023.

\bibitem{eer3}
S.~Ding, Y.~Zhang, and Z.~Duan, ``Samo: Speaker attractor multi-center one-class learning for voice anti-spoofing,'' \emph{ICASSP 2023 - 2023 IEEE International Conference on Acoustics, Speech and Signal Processing (ICASSP)}, pp. 1--5, 2022.

\bibitem{ASVspoof2A}
X.~Wang, J.~Yamagishi, M.~Todisco, H.~Delgado, A.~Nautsch, N.~W.~D. Evans, M.~Sahidullah, V.~Vestman, T.~H. Kinnunen, K.~A. LEE, L.~Juvela, P.~Alku, Y.-H. Peng, H.-T. Hwang, Y.~Tsao, H.-M. Wang, S.~L. Maguer, M.~Becker, and Z.~Ling, ``Asvspoof 2019: A large-scale public database of synthesized, converted and replayed speech,'' \emph{Comput. Speech Lang.}, vol.~64, p. 101114, 2019.

\bibitem{SingFakeSV}
Y.~Zang, Y.~Zhang, M.~Heydari, and Z.~Duan, ``Singfake: Singing voice deepfake detection,'' \emph{ICASSP 2024 - 2024 IEEE International Conference on Acoustics, Speech and Signal Processing (ICASSP)}, pp. 12\,156--12\,160, 2023.

\bibitem{chen24o_interspeech}
X.-B. Chen, H.~Wu, J.-S.~R. Jang, and H.~yi~Lee, ``Singing voice graph modeling for singfake detection,'' \emph{Interspeech 2024}, 2024.

\bibitem{chen2022unispeech}
S.~Chen, Y.~Wu, C.~Wang, Z.~Chen, Z.~Chen, S.~Liu, J.~Wu, Y.~Qian, F.~Wei, J.~Li, and X.~Yu, ``Unispeech-sat: Universal speech representation learning with speaker aware pre-training,'' \emph{ICASSP 2022 - 2022 IEEE International Conference on Acoustics, Speech and Signal Processing (ICASSP)}, pp. 6152--6156, 2021.

\bibitem{chen2022wavlm}
S.~Chen, C.~Wang, Z.~Chen, Y.~Wu, S.~Liu, Z.~Chen, J.~Li, N.~Kanda, T.~Yoshioka, X.~Xiao \emph{et~al.}, ``Wavlm: Large-scale self-supervised pre-training for full stack speech processing,'' \emph{IEEE Journal of Selected Topics in Signal Processing}, vol.~16, no.~6, pp. 1505--1518, 2022.

\bibitem{baevski2020wav2vec}
A.~Baevski, Y.~Zhou, A.~Mohamed, and M.~Auli, ``wav2vec 2.0: A framework for self-supervised learning of speech representations,'' \emph{Advances in neural information processing systems}, vol.~33, pp. 12\,449--12\,460, 2020.

\bibitem{pepino21_interspeech}
L.~Pepino, P.~E. Riera, and L.~Ferrer, ``Emotion recognition from speech using wav2vec 2.0 embeddings,'' \emph{ArXiv}, vol. abs/2104.03502, 2021.

\bibitem{kang2024experimental}
T.~Kang, S.~Han, S.~Choi, J.~Seo, S.~Chung, S.~Lee, S.~Oh, and I.-Y. Kwak, ``Experimental study: Enhancing voice spoofing detection models with wav2vec 2.0,'' \emph{arXiv preprint arXiv:2402.17127}, 2024.

\bibitem{8461375}
D.~Snyder, D.~Garcia-Romero, G.~Sell, D.~Povey, and S.~Khudanpur, ``X-vectors: Robust dnn embeddings for speaker recognition,'' \emph{2018 IEEE International Conference on Acoustics, Speech and Signal Processing (ICASSP)}, pp. 5329--5333, 2018.

\bibitem{Phukan2023TransformingTE}
O.~C. Phukan, A.~B. Buduru, and R.~Sharma, ``Transforming the embeddings: A lightweight technique for speech emotion recognition tasks,'' in \emph{Interspeech}, 2023.

\bibitem{fukumori2023investigating}
T.~Fukumori, T.~Ishida, and Y.~Yamashita, ``Investigating the effectiveness of speaker embeddings for shout intensity prediction,'' \emph{2023 Asia Pacific Signal and Information Processing Association Annual Summit and Conference (APSIPA ASC)}, pp. 1838--1842, 2023.

\bibitem{li2023mert}
Y.~Li, R.~Yuan, G.~Zhang, Y.~Ma, X.~Chen, H.~Yin, C.~Xiao, C.~Lin, A.~Ragni, E.~Benetos \emph{et~al.}, ``Mert: Acoustic music understanding model with large-scale self-supervised training,'' \emph{arXiv preprint arXiv:2306.00107}, 2023.

\bibitem{Music2VecAS}
Y.~Li, R.~Yuan, G.~Zhang, Y.~Ma, C.~Lin, X.~Chen, A.~Ragni, H.~Yin, Z.~Hu, H.~He, E.~Benetos, N.~Gyenge, R.~Liu, and J.~Fu, ``Map-music2vec: A simple and effective baseline for self-supervised music audio representation learning,'' \emph{ArXiv}, vol. abs/2212.02508, 2022.

\bibitem{CKA1}
S.~Kornblith, M.~Norouzi, H.~Lee, and G.~E. Hinton, ``Similarity of neural network representations revisited,'' \emph{ArXiv}, vol. abs/1905.00414, 2019.

\bibitem{CKA2}
M.~Raghu, T.~Unterthiner, S.~Kornblith, C.~Zhang, and A.~Dosovitskiy, ``Do vision transformers see like convolutional neural networks?'' in \emph{Neural Information Processing Systems}, 2021.

\bibitem{CKA3}
M.~Maniparambil, R.~Akshulakov, Y.~A.~D. Djilali, S.~Narayan, M.~E.~A. Seddik, K.~Mangalam, and N.~E. O'Connor, ``Do vision and language encoders represent the world similarly?'' \emph{ArXiv}, vol. abs/2401.05224, 2024.

\bibitem{CKA4}
T.~Nguyen, M.~Raghu, and S.~Kornblith, ``Do wide and deep networks learn the same things? uncovering how neural network representations vary with width and depth,'' \emph{ArXiv}, vol. abs/2010.15327, 2020.

\bibitem{CKA5}
M.-J. Davari, S.~Horoi, A.~Natik, G.~Lajoie, G.~Wolf, and E.~Belilovsky, ``Reliability of cka as a similarity measure in deep learning,'' \emph{ArXiv}, vol. abs/2210.16156, 2022.

\bibitem{CtrSVDDAB}
Y.~Zang, J.~Shi, Y.~Zhang, R.~Yamamoto, J.~Han, Y.~Tang, S.~Xu, W.~Zhao, J.~Guo, T.~Toda, and Z.~Duan, ``Ctrsvdd: A benchmark dataset and baseline analysis for controlled singing voice deepfake detection,'' \emph{ArXiv}, vol. abs/2406.02438, 2024.

\bibitem{FSDAI}
Y.~Xie, J.~Zhou, X.~Lu, Z.~Jiang, Y.~Yang, H.~Cheng, and L.~Ye, ``Fsd: An initial chinese dataset for fake song detection,'' \emph{ICASSP 2024 - 2024 IEEE International Conference on Acoustics, Speech and Signal Processing (ICASSP)}, pp. 4605--4609, 2023.

\end{thebibliography}

\clearpage

\end{document}